# Suppression of crosstalk in coupled plasmonic waveguides

E. V. Kuznetsov,[1,2,3] A. M. Merzlikin,[1,2,3] A. A. Zyablovsky,[1,2,3] A. P. Vinogradov,[1,2,3] and A. A. Lisyansky[4,5]

[1] *Dukhov Research Institute for Automatics, 22 Sushchevskaya, Moscow 127055, Russia*

[2]*Moscow Institute of Physics and Technology, 9 Institutskiy per., Dolgoprudniy 141700, Moscow Reg., Russia*

[3]*Institute for Theoretical and Applied Electromagnetics RAS, 13 Izhorskaya, Moscow 125412, Russia*

[4]*Department of Physics, Queens College of the City University of New York, Queens, NY 11367, USA*

[5]*The Graduate Center of the City University of New York, New York, New York 10016, USA*

We demonstrate the suppression of crosstalk between two dielectric nanowaveguides by placing an auxiliary linear waveguide between loaded waveguides spaced by one wavelength. The total cross-sectional dimension of the system containing two transmission lines is less than two microns that is hundred times smaller than a cross-section of a system made of dielectric fiber. The propagating modes in these waveguides are the sum and the difference of symmetric and antisymmetric modes of the coupled system. Crosstalk is suppressed by matching the wavenumbers of these modes. The analytically obtained results are confirmed by numerical simulation.

High-quality surface plasmonic waveguides are key building blocks of nanoplasmonic-based optical devices.[1-9] Among other applications, such waveguides could be used as chip-to-chip[10] and on-chip[11, 12] plasmonic interconnects. The advantage of surface plasmonic waveguides is their small size and high frequency which can reach optical frequencies. This is achieved by virtue of subwavelength localization of the electromagnetic field. The main shortcoming of plasmonic waveguides is their attenuation due to Ohmic losses and crosstalk between waveguides. The former may be compensated by using active media.[13-18] The latter arises due to signal tunneling from one waveguide to another. As in other waveguides, the surface plasmon wave leaks out beyond the physical boundary of the waveguide. As a result, when two waveguides are in close proximity, energy cross-flow between them causes crosstalk.

To achieve greater miniaturization and to increase the throughput of optical lines, it would be desirable to use a high density of plasmonic waveguides which inevitably leads to an increase in the crosstalk. In Ref. 19 an original method for the crosstalk suppression was proposed. An additional *nonlinear* waveguide is placed between the waveguides carrying the signal. By using adiabatic elimination, the authors of Ref. 19 demonstrated decoupling of linear waveguides experimentally. Even though the suggested approach allows for controlling waveguides, the necessity of using nonlinear media for tuning the propagation constant of the middle waveguide may present a major technical obstacle.

In the present paper, we study theoretically the possibility of decoupling nanoplasmonic waveguides by placing another waveguide between them. In contrast to the scheme suggested in Ref. 19, in our scheme, the additional waveguide is *linear*. The additional waveguide changes the dispersion relations of the modes. Its parameters are chosen so that in the carrying waveguides, the symmetric and antisymmetric modes have the same wavenumber. In this case, any linear combination of these modes is transmitted without changed. In particular, this applies to the excitation of such a linear combination when a signal is carried by a single waveguide.

Within the framework of the coupled mode theory, we consider the propagation of an electromagnetic wave in a system comprised of two coupled waveguides with the same wavenumbers, $\beta$. We assume that the $z$-axis is directed along the waveguides and denote the coupling constant between the waveguides as $\kappa$. The field in the waveguide is determined by the wave equation

$$i\frac{d}{dz}\begin{pmatrix} u_1(z) \\ u_2(z) \end{pmatrix} = \begin{pmatrix} \beta & \kappa \\ \kappa^* & \beta \end{pmatrix}\begin{pmatrix} u_1(z) \\ u_2(z) \end{pmatrix}, \tag{1}$$

where $u_1$ and $u_2$ are field amplitudes in the respective waveguides. The eigenmodes of the systems can be found as eigenvectors of the matrix $\begin{pmatrix} \beta & \kappa \\ \kappa^* & \beta \end{pmatrix}$:

$$\begin{pmatrix} u_1 \\ u_2 \end{pmatrix}_+ = \begin{pmatrix} \kappa/|\kappa| \\ 1 \end{pmatrix}, \quad \begin{pmatrix} u_1 \\ u_2 \end{pmatrix}_- = \begin{pmatrix} -\kappa/|\kappa| \\ 1 \end{pmatrix}, \tag{2}$$

and their wavenumbers are eigenvalues of the same matrix:

$$\beta_+ = \beta + |\kappa|, \quad \beta_- = \beta - |\kappa|, \tag{3}$$

One mode is reflection symmetric, and another is reflection antisymmetric. Therefore, if the phase difference between modes is zero ($\pi$), the field exists in the first (second) waveguide only.

The energy of the first waveguide is completely transferred into the second one when the phase difference between eigenmodes changes by $\pi$. This happens when the distance between waveguides satisfies the condition:

$$\mathrm{Re}\left(\beta_+ - \beta_-\right)L_{CT} = 2|\kappa|L_{CT} = \pi \ . \tag{4}$$

In the case of two waveguides, the only way to increase the crosstalk length, $L_{CT}$, is by decreasing $|\kappa|$. This can be done either by increasing the distance between waveguides, which decreases the throughput of the system per unit area or by restructuring the system.

Let us now discuss the effect of loss on crosstalk in the suggested system. In a system of two coupled lossy waveguides, wavenumbers of eigenmodes acquire imaginary parts:

$$\beta_+ = \beta + |\kappa| + i\gamma_+, \quad \beta_- = \beta - |\kappa| + i\gamma_-. \tag{5}$$

If loss is the same in both waveguides, then the imaginary parts of symmetric and antisymmetric modes are also the same $\gamma_+ = \gamma_-$. As a result, the crosstalk length $L_{CT} = \pi / \left(2\operatorname{Re}\left(\beta_+ - \beta_-\right)\right)$ does not depend on loss.

If imaginary parts of symmetric and antisymmetric modes are different, then when the wave propagates along one of the waveguides, the signal leaks into the second waveguide. As a result, there is a mixture of signals in the waveguides having a part that oscillates at the $L_{CT} = \pi / \left(2\operatorname{Re}\left(\beta_+ - \beta_-\right)\right)$ length and a decaying part, $0.5\left(1 - exp\left(-\Delta\gamma L\right)\right)$. Thus, an effective crosstalk length becomes $min\left\{L_{CT}, \Delta\gamma^{-1}\right\}$. The scheme suggested in this paper is applicable when $L_{CT} << \Delta\gamma^{-1}$.

To increase $L_{CT}$ without increasing the distance between waveguides, we place an auxiliary *linear* waveguide between waveguides carrying a signal (see Fig. 1). We assume that carrying waveguides have the same wavenumbers $\beta_1$, and the coupling constant between them is $\kappa_{12}$. The wavenumber of the auxiliary waveguide is $\beta_2$ and the coupling constant between this waveguide and carrying waveguides is $\kappa_{13}$.

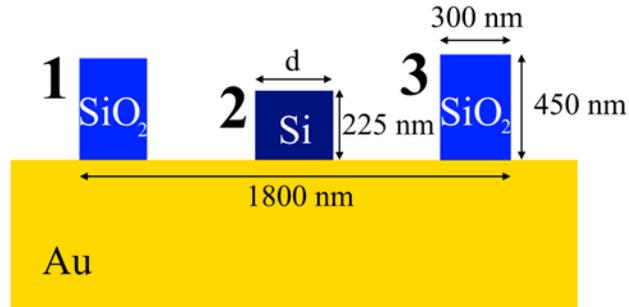

FIG. 1. The schematics of the surface plasmon waveguide system. The auxiliary waveguide (2) is placed between two carrying signal waveguides, (1) and (3). All waveguides are interfaced with a metal substrate. The width of the central waveguide is varied.

Let us consider this system of waveguides in the approximation of coupled modes. Assuming that the waveguides are in the $xz$-plane and the $z$-axis is directed along the waveguides we can write the equation that determines the coordinate dependence of the field:

$$i\frac{d}{dz}\begin{pmatrix} u_1(z) \\ u_2(z) \\ u_3(z) \end{pmatrix} = \begin{pmatrix} \beta_1 & \kappa_{12} & \kappa_{13} \\ \kappa_{12}^* & \beta_2 & \kappa_{12} \\ \kappa_{13}^* & \kappa_{12}^* & \beta_1 \end{pmatrix}\begin{pmatrix} u_1(z) \\ u_2(z) \\ u_3(z) \end{pmatrix},$$ (6)

where $u_1$, $u_2$, and $u_3$ are field amplitudes in the respective waveguides. Two of the eigenmodes of this system are symmetric, and the third one is antisymmetric. The field of the latter mode is zero in the central waveguide. Fields of symmetric modes are nonzero in all waveguides. However, one symmetric mode has a maximum in the central waveguide, while the other has maxima in the side waveguides. Our goal is to find parameters of waveguides in such a way that wavenumbers of the antisymmetric mode $U_-$ and the symmetric mode having its maximum in the side waveguides $U_{+1}$ coincide. Then, according to Eq. (4), $L_{CT} = \infty$.

For our system, wavenumbers $\beta_{+1}$, $\beta_{+2}$, and $\beta_-$ can be expressed as

$$\beta_{+1} = \frac{1}{2}\left(\beta_1 + \beta_2 + \kappa_{13} - \sqrt{\beta_1^2 - 2\beta_1\beta_2 + \beta_2^2 + 8\kappa_{12}^2 + 2\beta_1\kappa_{13} - 2\beta_2\kappa_{13} + \kappa_{13}^2}\right),$$ (7)

$$\beta_{+2} = \frac{1}{2}\left(\beta_1 + \beta_2 + \kappa_{13} + \sqrt{\beta_1^2 - 2\beta_1\beta_2 + \beta_2^2 + 8\kappa_{12}^2 + 2\beta_1\kappa_{13} - 2\beta_2\kappa_{13} + \kappa_{13}^2}\right),$$ (8)

$$\beta_- = \beta_1 - \kappa_{13}. \tag{9}$$

Below we normalize all wave numbers by the wavenumber of the free space $\beta_0 = \omega / c$. At the first step, to estimate the possibility of the crosstalk abatement, we use parameters from the range that is observed in experiments.[20, 21] We assume that the wavenumbers of the side waveguides are $\beta_1 = \beta_3 = 2$. The coupling constants between waveguides $\kappa_{12}$ and $\kappa_{13}$ falloff exponentially with the distance between them. The coupling constant between the first and the second (or the second and the third) waveguides $\kappa_{12}$ should be a few times larger than $\kappa_{13}$. The coupled mode theory that we use is a perturbation theory with respect to the coupling constants $\kappa_{12}$ and $\kappa_{13}$. The theory is applicable if $\kappa_{12}, \kappa_{13} \ll \beta_1$. We set $\kappa_{12} = 10^{-2} \beta_1 = 2 \cdot 10^{-2}$ and $\kappa_{13} = \kappa_{12} / 4 = 5 \cdot 10^{-3}$, while the wavenumber $\beta_2$ of the central waveguide varies. For these parameters, in the absence of the central waveguide, the crosstalk length is $L_{CT} = \pi / 2 |\kappa_{13}| = 50 \lambda_\sigma$. The crosstalk length limits the length of a waveguide at which the information transmission is possible. In our example, the waveguide length cannot exceed $50 \lambda_\sigma$, because, at this distance, the signal is completely transferred from the first to the third waveguide.

The dependencies of wavenumbers of each mode on $\beta_2$ are shown in Fig. 2. One can see that for $\beta_2 = 2.1102$, the wavenumbers for the modes $U_-$ and $U_{+1}$ coincide and are equal to 1.995. At this point, eigenmodes of the system are

$$U_{+1} = \begin{pmatrix} 0.68 \\ 0.27 \\ 0.68 \end{pmatrix}, \quad U_{+2} = \begin{pmatrix} -0.19 \\ -0.96 \\ -0.19 \end{pmatrix}, \quad U_- = \begin{pmatrix} 0.707 \\ 0 \\ -0.707 \end{pmatrix}. \tag{10}$$

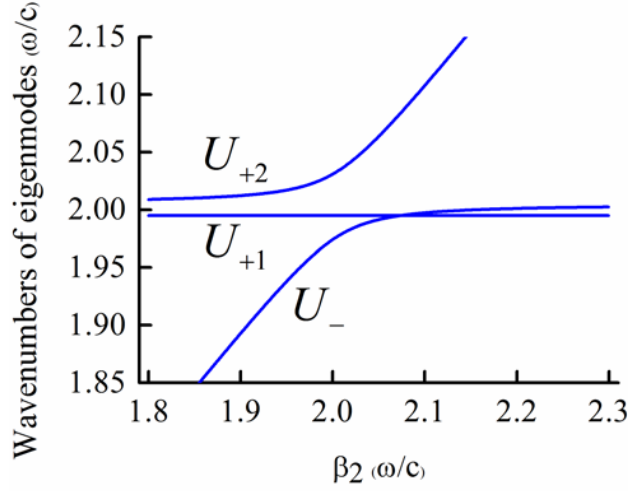

FIG. 2. Dependencies of wavenumbers of eigenmodes of the system comprised of three coupled waveguides on the wavenumber in the central waveguide.

Let us find the field distribution in the waveguides assuming that a signal with the unit amplitude is initiated in the first waveguide only. In this case, the amplitudes of the eigenmodes are

$$c_{+1} = U_{+1}(1) = 0.68, \quad c_{+2} = U_{+2}(1) = -0.19, \quad c_{-} = U_{-}(1) = 0.707 .$$ (11)

Since wavenumbers of eigenmodes $U_{-}$ and $U_{+1}$ are the same, the phase difference between these modes does not change. In the first waveguide, the field can be found as

$$\left| u_1(z) \right| = \left| c_{+1} U_{+1}(1) + c_{-} U_{-}(1) + \exp\left( i\left( \beta_{+2} - \beta_{+1} \right) z \right) c_{+2} U_{+2}(1) \right| .$$ (12)

This field reaches minimum values at the points $z_{\min} = (2n+1)\pi / \left( \beta_{+2} - \beta_{+1} \right)$, where $n$ is an integer. The minimum value of $u_1$ at these points is

$$\left| u_1(z) \right| = \left| c_{+1} U_{+1}(1) + c_{-} U_{-}(1) - c_{+2} U_{+2}(1) \right| = \left| 1 - 2 c_{+2} U_{+2}(1) \right| \approx 0.92 .$$ (13)

In the third waveguide, the field amplitude is

$$\left|u_2(z)\right| = \left|c_{+1}U_{+1}(3) + c_-U_-(3) + \exp\left(i\left(\beta_{+2} - \beta_{+1}\right)z\right)c_{+2}U_{+2}(3)\right|. \qquad (14)$$

This amplitude has maxima at the same points, $z_{\min}$, at which $u_1$ is minimal. The value of $u_2$ at these points is

$$\left|u_2(z)\right| = \left|c_{+1}U_{+1}(3) + c_-U_-(3) - c_{+2}U_{+2}(3)\right| = \left|2c_{+2}U_{+2}(3)\right| \approx 0.07. \qquad (15)$$

The maximum value of the field intensity in the third waveguide is $\approx 5 \cdot 10^{-3}$ of the field intensity in the first waveguide. Thus, in our system, the intensity of the induced field is on the level of $0.5\%$ of the intensity of the carrying signal. In other words, in the system, we constructed crosstalks between waveguides are practically suppressed.

In a system of three waveguides, the situation is similar that of two waveguides: if the loss is the same in all waveguides, the crosstalk length is not affected. Indeed, for the wavenumbers $\beta_1 \to \beta_1 + i\gamma$ и $\beta_2 \to \beta_2 + i\gamma$ Eq. (5) has the form:

$$i\frac{d}{dz}\begin{pmatrix} u_1(z) \\ u_2(z) \\ u_3(z) \end{pmatrix} = \begin{pmatrix} \beta_1 & \kappa_{12} & \kappa_{13} \\ \kappa_{12}^* & \beta_2 & \kappa_{12} \\ \kappa_{13}^* & \kappa_{12}^* & \beta_1 \end{pmatrix}\begin{pmatrix} u_1(z) \\ u_2(z) \\ u_3(z) \end{pmatrix} + i\begin{pmatrix} \gamma & 0 & 0 \\ 0 & \gamma & 0 \\ 0 & 0 & \gamma \end{pmatrix}\begin{pmatrix} u_1(z) \\ u_2(z) \\ u_3(z) \end{pmatrix}. \qquad (16)$$

The new variables $u_i(z)' = u_i(z)\exp(-\gamma z)$ satisfy Eq. (5). For these variables, the same arguments as for systems with no loss can be applied. Generally, in the auxiliary and signal carrying waveguides, losses are different. However, the parameters of the auxiliary waveguide can be tuned in a way that loss in the auxiliary and carrying signal waveguides are close, as shown in the example below. If the loss is different in all waveguides, then the situation is similar to that in two waveguides: the transmission length of a signal is defined by the smaller of the damping and crosstalk lengths.

In a system of waveguides, within the framework of the theory of coupled modes, the solution is represented as a linear combination of eigenmodes of each waveguide. This theory is the first-order perturbation theory in which the perturbation parameters are coupling constants between the waveguides. Higher orders of the perturbation theory may substantially change the result obtained above because even a small difference between wavenumbers of symmetric and antisymmetric modes may result in an energy flux between waveguides. To verify the analytical results, in the next section, we perform numerical simulation of the mode propagation in the waveguide system described above.

Let us consider a system of dielectric-loaded surface plasmonic waveguides[22, 23] with parameters shown in Fig. 1. The waveguides are dielectric strips on a metal substrate which is modeled by metal (Au) film of thickness 300 nm. At the telecom wavelength $\lambda = 1.55\,\mu m$, the refractive index of gold is $0.5241 + 10.742i$.[24]

Let us assume that the side waveguides have the same refractive indices 1.5277 (SiO$_2$).[25] This warrants a single-mode regime of each waveguide. The index of refraction of the auxiliary central waveguide is 3.4757 (Si).[26] The widths of side waveguides are fixed; the width of the central waveguide may vary.

We obtain eigenmodes of the system by performing finite-element-method calculations for a mesh size varying from 52 nm to 260 nm. Closer to the system boundary, the cell size is smaller, farther away from the boundary it is larger. The width of the calculation range was chosen as 6 μm and its height as 3.3 μm. Thus, the system size is much smaller than the calculation region.

Numerical simulation shows that the equality of the wavevectors for symmetric and antisymmetric modes having maxima of intensities in the side waveguides is reached for the width of the central waveguide of $d = 175$ nm. There are three eigenmodes in the system (Fig. 3). For $d = 175$ nm, the real part of wavenumbers of symmetric and antisymmetric modes with maxima at the side waveguides (Figs. 3a and 3b) are the same; their wavenumbers are $\beta_1 = 1.0856 + 0.00252i$ and $\beta_2 = 1.0856 + 0.00271i$ for the wavelength 1.55 µm. The third mode has maxima at the central waveguide (Fig. 3c).

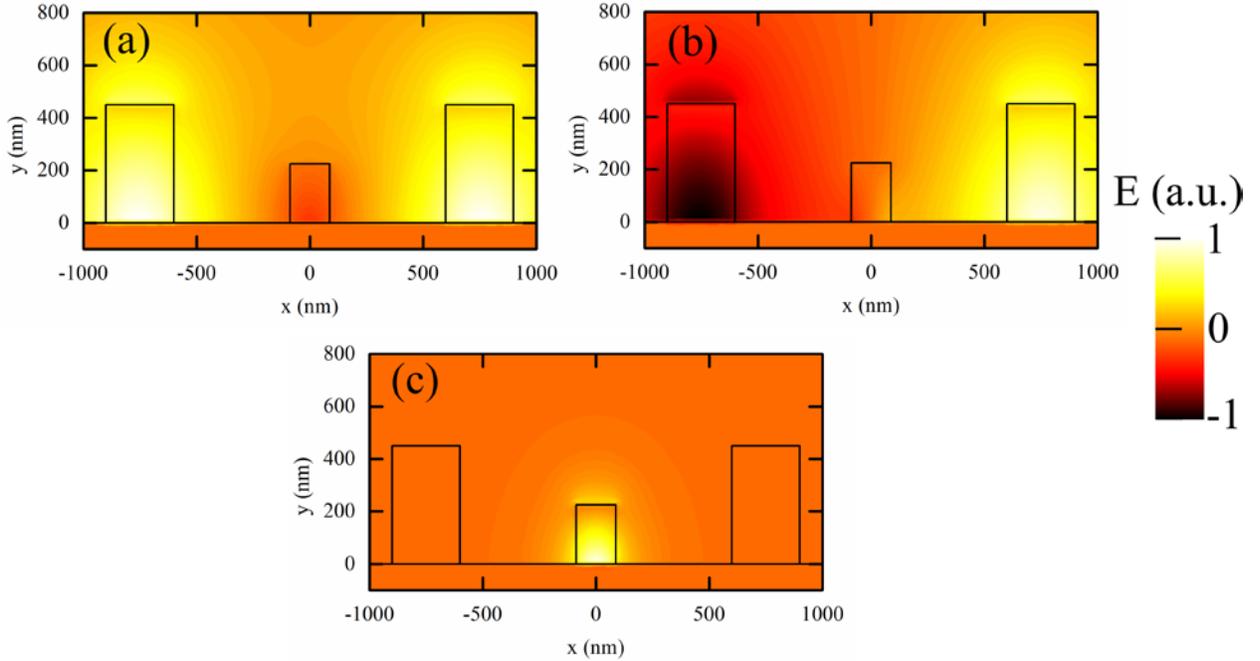

FIG. 3. The screenshot of the electric field distribution of the eigenmodes: (a) $\beta_1 = 1.0856 + 0.0025i$, (b) $\beta_2 = 1.0856 + 0.0027i$, (c) $\beta_3 = 2.4636 + 0.0288i$.

Let us now consider the propagation of the signal incident in the third waveguide in our system. We assume that the lengths of the first and second waveguides are 30 µm each, while the length of the third, signal carrying, waveguide is 32 µm. Thus an excitation of this waveguide is performed on the additional piece AB of 2 µm in length (see inset in Fig. 4). The other parameters are the same as noted above. The schematics of the system used in our numerical simulation is shown in the inset of Fig. 4. The signal with the same field distribution as in the eigenmode of the waveguide is incident on the 32-µm-waveguide.

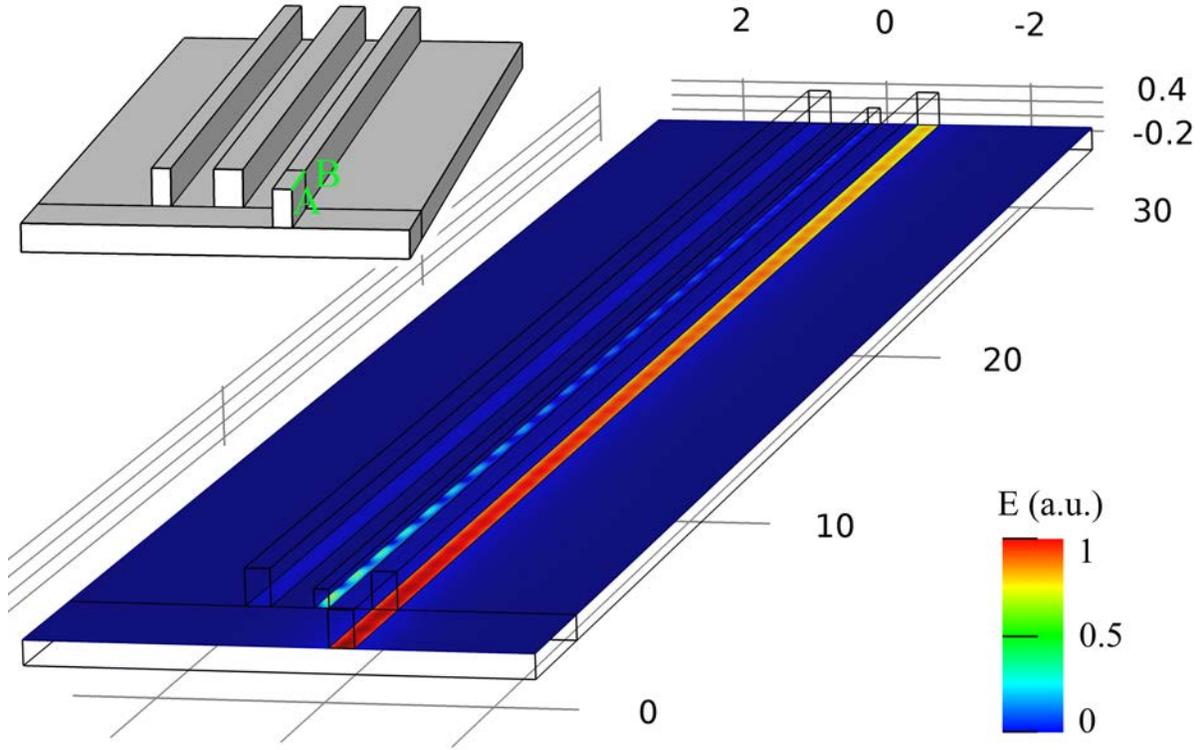

FIG. 4. The distribution of the absolute value of the electric field on the interface of the Au-substrate and three waveguides. Inset: the schematics of the system. The interval AB of the third waveguide is used to excite this waveguide.

Numerical simulation shows that the signal propagates along the right waveguide with a very small leakage into the left waveguide. The fields in each waveguide are shown in Fig. 5a.

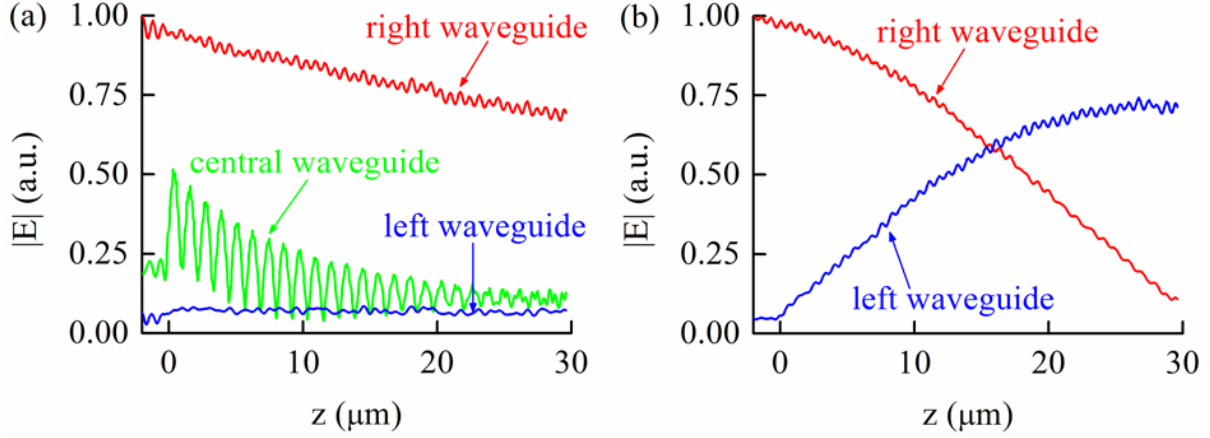

FIG. 5. The distribution of the absolute values of the electric fields at the center of the interface between the Au-substrate and the right (red), central (green), and left (blue) waveguides with (a) and without (b) the auxiliary waveguide. The origin of $z$-axis is point B in Fig. 4.

As one can see from Fig. 5a, not the whole energy is transferred along the signal-carrying third waveguide. This is mainly due to Ohmic losses in metal. Another reason is that the mode excited at the section AB of the signal-carrying waveguide does not match exactly to the sum of symmetric and antisymmetric modes of the system of three waveguides. At the boundary of the transition from one waveguide (AB) to three waveguides, some part of the energy is used to excite modes at the first and second waveguides.

In order to demonstrate the contribution of the central waveguide to the crosstalk suppression, we have also simulated the system of plasmonic waveguides without the auxiliary central waveguide (see Fig. 6). One can see that as the wave propagates along the left waveguide, its energy is transferred to the other waveguide (see Fig. 5b and Fig. 6).

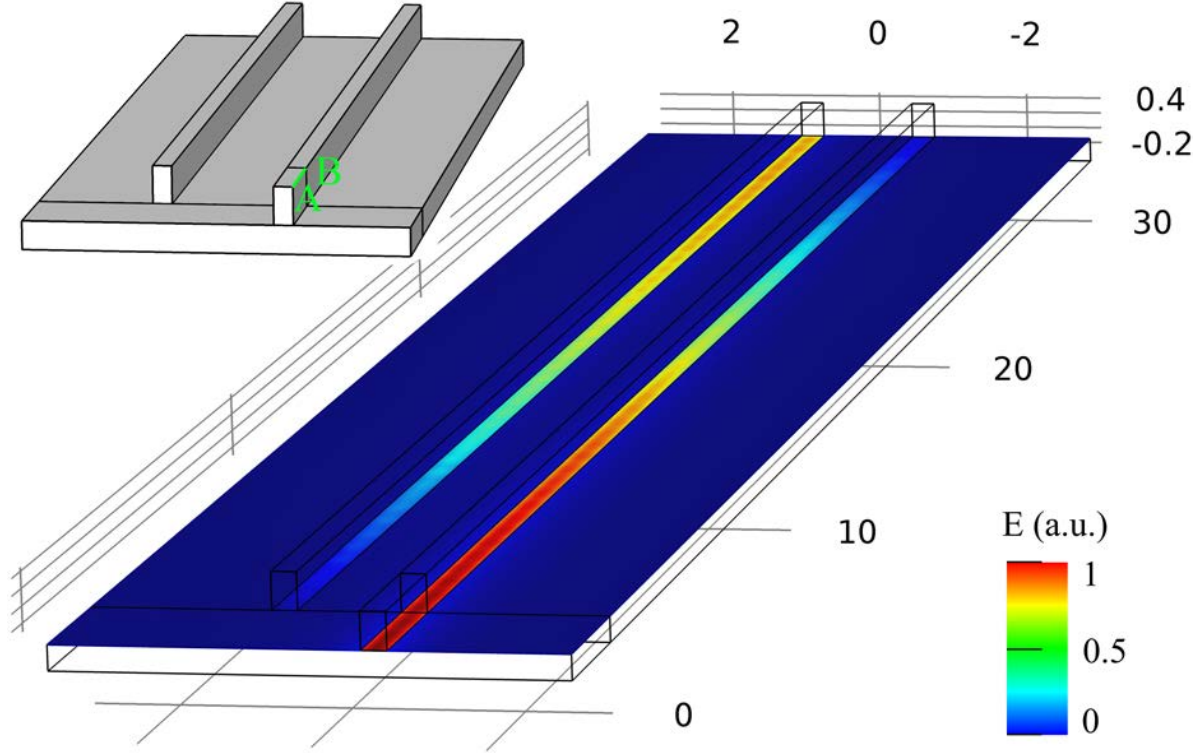

FIG. 6. The distribution of the absolute value of the electric field in the system shown in Fig. 4 but without the auxiliary waveguide. Inset: the schematics of the system.

To conclude, we show analytically and numerically that the energy flux between two waveguides can be suppressed by placing an auxiliary waveguide with specially chosen parameters between them. The wave excited in one of the waveguides, which is the sum of symmetric and antisymmetric modes, corresponds to a zero field in the other two waveguides. Since the wavenumbers of symmetric and antisymmetric modes are the same, this relationship does not change as the wave propagates along the system The suggested set-up allows for a substantial increase in the crosstalk length between surface plasmonic waveguides.

A.A.L would like to acknowledge support from the NSF under Grant No. DMR-1312707.

## References

[1] L. Dobrzynski and A. A. Maradudin, Phys. Rev. B **6**, 3810 (1972).

[2] L. C. Davis, Phys. Rev. B **14**, 5523 (1976).

[3] A. Eguiluz and A. A. Maradudin, Phys. Rev. B **14**, 5526 (1976).

[4] A. D. Boardman, Phys. Rev. B **24**, 5703 (1981).

[5] A. D. Boardman, Phys. Rev. B **32**, 6045 (1985).

[6] J. Q. Lu and A. A. Maradudin, Phys. Rev. B **42**, 11159 (1990).

[7] I. V. Novikov and A. A. Maradudin, Phys. Rev. B **66**, 035403 (2002).

[8] D. F. P. Pile and D. K. Gramotnev, Opt. Lett. **29**, 1069 (2004).

[9] S. I. Bozhevolnyi, V. S. Volkov, E. Devaux, J.-Y. Laluet, and T. W. Ebbesen, Nature **440**, 508 (2006).

[10] J. T. Kim, J. J. Ju, S. Park, M.-S. Kim, S. K. Park, and M.-H. Lee, Opt. Express **16**, 13133 (2008).

[11] V. J. Sorger, R. F. Oulton, R.-M. Ma, and X. Zhang, MRS Bulletin **37**, 008 (2012).

[12] J. A. Conway, S. Sahni, and T. Szkopek, Opt. Express **15**, 4474 (2007).

[13] J. Grandidier, G. C. des Francs, S. Massenot, A. Bouhelier, L. Markey, J.-C. Weeber, C. Finot, and A. Dereux, Nano Lett. **9**, 2935 (2009).

[14] M. T. Hill, M. Marell, E. S. P. Leong, B. Smalbrugge, Y. Zhu, M. Sun, P. J. van Veldhoven, E. J. Geluk, F. Karouta, Y.-S. Oei, R. Nötzel, C.-Z. Ning, and M. K. Smit, Opt. Express **17**, 11107 (2009).

[15] P. Berini and I. de Leon, Nat. Photon **6**, 16 (2012).

[16] R. F. Oulton, V. J. Sorger, T. Zentgraf, R.-M. Ma, C. Gladden, L. Dai, G. Bartal, and X. Zhang, Nature **461**, 629 (2009).

[17] I. de Leon and P. Berini, Nat. Photon **4**, 382 (2010).

[18] A. A. Lisyansky, I. A. Nechepurenko, A. V. Dorofeenko, A. P. Vinogradov, and A. A. Pukhov, Phys. Rev. B **84**, 153409 (2011).

[19] M. Mrejen, H. Suchowski, T. Hatakeyama, C. Wu, L. Feng, K. O'Brien, Y. Wang, and X. Zhang, Nat. Commun. **6**, 7565 (2015).

[20] C. E. Rüter, K. G. Makris, R. El-Ganainy, D. N. Christodoulides, M. Segev, and D. Kip, Nature Phys. **6**, 192 (2010).

[21] A. Guo, G. J. Salamo, D. Duchesne, R. Morandotti, M. Volatier-Ravat, V. Aimez, G. A. Siviloglou, and D. N. Christodoulides, Phys. Rev. Lett. **103**, 093902 (2009).

[22] T. Holmgaard and S. I. Bozhevolnyi, Phys. Rev. B **75**, 245405 (2007).

[23] T. Holmgaard, S. I. Bozhevolnyi, L. Markey, and A. Dereux, Appl. Phys. Lett. **92**, 011124 (2008).

[24] P. B. Johnson and R. W. Christy, Physical Review B **6**, 4370 (1972).

[25] I. H. Malitson, J. Opt. Soc. Am. **55**, 1205 (1965).

[26] H. H. Li, J. Phys. Chem. Ref. Data **9**, 561 (1993).